\documentclass[conference]{IEEEtran}
\IEEEoverridecommandlockouts
\usepackage{algorithm}
\usepackage{subcaption}
\usepackage{multirow}
\usepackage[noend]{algpseudocode}
\usepackage{cite}
\usepackage{amsmath,amssymb,amsfonts}
\usepackage{graphicx}
\usepackage{textcomp}
\usepackage{xcolor}
\usepackage{algorithm}
\usepackage[noend]{algpseudocode}
\usepackage{algorithmicx}
\usepackage{booktabs}
\usepackage{bbm}
\usepackage{comment}
\usepackage{amsthm}
\newtheorem{definition}{Definition}
\usepackage{tikz}
\usetikzlibrary{positioning,fit,backgrounds,arrows.meta,calc}
\usepackage{soul}
\usepackage{url}
\usepackage[hidelinks]{hyperref}
\usepackage{xurl}

\definecolor{cpubg}{RGB}{228,238,248}
\definecolor{cpublk}{RGB}{183,208,232}
\definecolor{gpubg}{RGB}{230,242,232}
\definecolor{gpublk}{RGB}{184,214,190}
\definecolor{memblk}{RGB}{246,246,244}
\usepackage{tikz}
\newcommand*\circled[1]{\tikz[baseline=(char.base)]{
  \node[shape=circle,draw,inner sep=1pt] (char) {\footnotesize #1};}}
\usetikzlibrary{positioning,fit,backgrounds,arrows.meta}
\def\BibTeX{{\rm B\kern-.05em{\sc i\kern-.025em b}\kern-.08em
    T\kern-.1667em\lower.7ex\hbox{E}\kern-.125emX}}
\begin{document}

\title{BOA: Beamwidth Online Adaptation for 
Filtered-ANNS on a GPU\\
}

\author{\IEEEauthorblockN{Farhana Akter Tumpa}
\IEEEauthorblockA{\textit{Department of Computer Science} \\
\textit{University of California}\\
Riverside, USA \\
ftump001@ucr.edu}
\and
\IEEEauthorblockN{Rajiv Gupta}
\IEEEauthorblockA{\textit{Department of Computer Science} \\
\textit{University of California}\\
Riverside, USA \\
rajivg@ucr.edu}
}

\maketitle

\begin{abstract}
Filtered approximate nearest neighbor search, i.e. returning the top-$k$ vectors nearest to a query vector among those satisfying one or more attribute predicates, has become a fundamental operation in modern vector search systems. 
Graph-based solutions employ beam search to solve a batch of queries in parallel for high throughput and employ high \emph{fixed beamwidth} of 100 or greater for ensuring high recall. We observe that, given a batch of queries, more than half of the queries across multiple data sets can be solved precisely with a beamwidth of just 50 or less. Therefore, existing systems based on fixed high beamwidth sacrifice throughput to achieve high recall by forcing every query to search as thoroughly as the hardest query in the batch even though majority of queries can be resolved by a shallow search. 

In this paper we present a filtered ANNS engine for a single GPU named BOA that uses online beamwidth adaptation to customize the search effort across queries within a batch under \emph{multi-attribute} range filters. We address the recall throughput tradeoff with a multi-phase search: all queries are first evaluated under a narrow beam, and only those with uncertain results are progressively refined with wider beamwidths. This renders recall largely insensitive to the starting beamwidth, whereas prior methods must use a fixed high beamwidth for high recall. BOA+ overlaps execution of phases to further enhance throughput. 
Our experiments show that, for 10,000 queries, online adaptation achieves 94.05\% to 99.96\% recall with average beamwidth ranging from 22 to 77, while a non-adaptive approach requires a fixed beamwidth of 500 to achieve similar or lower recall. Consequently, adaptivity increases throughput by 7$\times$ to 12.5$\times$.
\end{abstract}

\begin{IEEEkeywords}
nearest-neighbor search, filtering attributes, beam-width, recall, throughput, GPU.
\end{IEEEkeywords}

\section{Introduction}
\label{sec:intro}
The Approximate Nearest Neighbor Search (ANNS) on high-dimensional vectors is a core operation in modern retrieval~\cite{b1}, recommendation~\cite{b2,b3,b4} and
retrieval-augmented generation~\cite{b5,b6,b7} systems. In practice, a query usually comes with a condition attached, not just a vector.. Each vector carries structured attributes such as price, timestamp, category, tags, access permissions, and a query specifies not only a target vector but also a predicate that those attributes must satisfy such as, a value to match, an interval to fall within, and a set of tags to contain. Given a query vector and a predicate, the task is to return the \emph{top-k} points nearest to the query among those satisfying the predicate. Because embeddings are increasingly stored alongside structured metadata, filtered queries are now common~\cite{b8,b9}, and an index that can enforce the predicate is of immense practical use.

\begin{table}[!t]
    \caption{ANNS Methods.}
    \label{intro}
    \centering
    \renewcommand{\arraystretch}{1.35}
    \begin{tabular}{|c||c|c|c|c|} \hline
    \textsf{System} &\!\!\textsf{Beamwidth (L)}\!\!& \textsf{\#Attribute Filters (m)} &  \!\!\textsf{Platform}\!\! \\ \hline \hline
    \textsf{BANG}~\cite{b16} & \textsf{Fixed L} & \textsf{None (m=0)} & \textsf{GPU} \\
    \textsf{JAG}~\cite{b18} & \textsf{Fixed L} & \textsf{Single (m=1)} & \textsf{CPU} \\
    \!\!\textsf{Garfield}~\cite{b17}\!\! & \textsf{Fixed L} & \textsf{Multiple (m$\geq$1)} & \textsf{GPU} \\ \hline
    \!\!\textbf{\textsf{BOA}}\!\! & \textbf{\textsf{Adaptive L}} & \!\!\textbf{\textsf{Multiple (m$\geq$1)}}\!\!  & \textbf{\textsf{GPU}} \\ \hline
    \end{tabular}
    \vspace{-0.15in}
\end{table}

Graph-based indices are the state of the art for ANNS~\cite{b10,b12}. They connect each point to a small set of near neighbors and answer a query by greedily walking the graph toward the query vector, visiting only a small fraction of query relevant data. This traversal is regular enough to parallelize, and a line of work has mapped it onto GPUs~\cite{b13,b14,b15} where a large batch of queries are processed concurrently to achieve large throughput gains~\cite{b16}. Of the GPU-based systems available, BANG~\cite{b16} delivers high-throughput but only supports \emph{unfiltered} queries, while Garfield~\cite{b17} supports filtering via \emph{range} predicates. Garfield index partitions the attribute
space into intervals and employs a structure specific to ranges. Filtered search over a single graph has been addressed on the CPU by JAG~\cite{b18} but for a single attribute per query. Table~\ref{intro} compares these systems with ours.

\smallskip
\underline{\sf Fixed Beamwidth Challenge} Both GPU-based systems, BANG~\cite{b16} and Garfield~\cite{b17}, share a common limitation. The graph search is governed by a single parameter, the beamwidth $L$ which is the number of candidates kept active as the walk proceeds. The parameter leads to a direct recall-throughput tradeoff. A small $L$ searches little and is fast, sustaining high throughput but yielding low recall; a large $L$ searches thoroughly and attains high recall, but at low throughput. In existing GPU systems this width is fixed for the entire query batch, so recall can be raised only by widening the beam for every query and paying the throughput cost across all of them; thus, incapable of avoiding recall-throughput tradeoff. 

This tradeoff, however, is an artifact of searching every query identically rather than a property of the queries. Searched with a small $L$, most queries already retrieve
their true neighbors; their answers are settled and do not improve with a wider search. Only a minority, whose neighborhoods lie in sparse or heavily filtered
regions, fall short and genuinely need greater effort. A uniformly large $L$ pays the full cost on every query to serve this minority, spending on the easy majority a great deal of work that changes nothing. The difficulty of a query is not known in advance but revealed online by the search itself and a system that reads this signal need not pay the cost for hardest queries on all queries.

\smallskip
\underline{\sf Filtering Challenge} A filter is a hard constraint imposed on a graph whose edges are unaware of the constraint because the graph is built on vector proximity alone. Therefore, a vector may not be returned as a neighbor, no matter how close it is to the query vector, unless its attributes satisfy the predicate. Thus, finding a solution to a query is more difficult when only a small fraction of points satisfy the predicate. Under a given predicate, the valid neighbors are sparse and scattered, with few edges among them, so a greedy walk following vector-proximity edges spends nearly all its steps among invalid nodes and often halts at a local optimum containing no valid points, never reaching the neighbors it was meant to find. Discarding invalid nodes mid-walk only makes this worse, since valid points are frequently reachable only through invalid intermediaries, and removing them disconnects the graph. Filtered search therefore cannot be recovered by constraining a vector-only graph at query time: the attribute structure must be built into the index, and the traversal must remain aware of the filter as it walks.


\smallskip
\paragraph*{\bf Our Approach} Our system exploits the observation that all queries in a batch do not require an equally high beamwidth to achieve high recall. In fact, we observe that more than half the queries across multiple data sets can be solved with a beamwidth of just 50 or less.
Therefore, we vary the search effort across queries within a batch instead of fixing the beamwidth for all of them, 
Instead of using a fixed beamwidth, the search runs in  phases: every query is first searched with a small $L$; each query's own result is then examined by how sharply its nearest candidates separate from the rest,  minority whose answers are not yet reliable; and only those queries are re-searched with a larger $L$. The easy majority is answered cheaply in fewer phases, the hard minority receives the thorough search it needs via more phases to deliver the high recall. While the existing GPU systems must trade recall for throughput, our recall is \emph{insensitive to the starting beamwidth $L$}. We handle the challenge posed by multi-attribute range filters on a single GPU as follows. Our algorithm searches over a single \emph{attribute-unaware} graph~\cite{b19} guided by a \emph{per-query filter signal} via a GPU beam search: candidates that violate the predicate are demoted in the search ordering rather than removed, so the walk is steered toward valid regions while those regions stay reachable through the demoted nodes. 
\textbf{\emph{Our BOA is the only system in Table~\ref{intro} that adapts beamwidth online and performs multi-attribute filtering on a GPU.}}

\smallskip
We make the following contributions:

\smallskip
\begin{itemize}\itemsep1pt
  \item \textsf{High Recall Without Wasteful Computation.} We introduce a multi-phase search that runs every query starting at a small $L$, identifies from their results the minority whose answers are not yet reliable, and searches only those with a larger $L$, decoupling recall from the throughput penalty that a uniformly wide beam imposes on every query. That is, via online adaptation of $L$, we ensure high recall while avoiding wasteful computation for high throughput.

  \item \textsf{Multi-Attribute Filtered Search on a GPU.} Our nearest neighbor engine supports multiple range filters and adapts traversal to them by carrying a filter signal through a GPU beam search so that filter satisfaction guides navigation while graph connectivity is preserved even in presence of multi-attribute filtering.

  \item \textsf{Evaluation.} We evaluate our approach on filtered-search queries and show that our method holds recall steady (94.05\% to 99.96\%) with low average beamwidth (22 to 77) while a non-adaptive approach requires very high fixed beamwidth (500) to attain similar or lower recall. Consequently, online beamwidth adaptation leads to 7$\times$ to 12.5$\times$ increase in throughput.
\end{itemize}

\section{Background \& Motivation}

\subsection{Background: Problem Definition}
\label{subsec:problem}

\noindent\textbf{Filtered approximate nearest neighbor search:}
Let $\mathcal{D} = \{p_1, \dots, p_n\}$ be a dataset of $n$ points, each a
pair $p = (x_p, a_p)$ with a vector $x_p \in \mathbb{R}^d$ and an
attribute record $a_p \in \mathcal{A}$. A query is a pair
$(q_i, f_i)$ with a query vector $q_i \in \mathbb{R}^d$ and a filter
$f_i \in \mathcal{F}$. A point is eligible only if its attributes satisfy
the filter, expressed by a binary matching function
$g : \mathcal{A} \times \mathcal{F} \rightarrow \{0,1\}$, and the eligible
points form the \emph{valid set}
$\mathcal{D}_f = \{\, p \in \mathcal{D} : g(a_p, f_i) = 1 \,\}$.
\begin{definition}Filtered $k$-NNS -
\label{def:fknn}
Given a dataset $\mathcal{D}$, a query $(q_i, f_i)$, and an integer
$k \le |\mathcal{D}_f|$, filtered $k$-nearest neighbor search returns the
set $\mathcal{K}_f \subseteq \mathcal{D}_f$ of $k$ points st:
\begin{equation}
\max_{p \in \mathcal{K}_f} \|x_p, q_i\| \;\le\;
\min_{p \in \mathcal{D}_f \setminus \mathcal{K}_f} \|x_p, q_i\|,
\label{eq:fknn}
\end{equation}
where $\|x_p, q_i\|$ denotes the distance between two vectors, either the
Euclidean ($L_2$) norm or cosine distance. When $|\mathcal{D}_f| < k$, the
answer is $\mathcal{D}_f$.
\end{definition}

We use the Euclidean norm throughout. Filter satisfaction is a hard
constraint, and distance ranks only among the points satisfying it;
setting $g \equiv 1$ recovers ordinary $k$-NNS. Computing $\mathcal{K}_f$
exactly requires a distance computation against every valid point, which
is prohibitive at scale, so practical systems solve the approximate
variant~\cite{b20,b21}. 

\smallskip
\noindent\textbf{Range-filtered nearest neighbor search:}
Vectors in production systems usually arrive with numeric metadata attached like a price, a timestamp, a rating, a geographic distance, and queries constrain that
metadata alongside the similarity search. A user browsing restaurants wants results that match a textual description, cost under some amount, and sit within
a few kilometers. Encoding the description as a vector reduces this to similarity search restricted to a numeric region, the problem known as \emph{range-filtered} nearest neighbor search (RFNNS), and this is the problem the paper addresses. Formally, the range-filtered setting fixes $\mathcal{A} = \mathbb{R}^m$: each
point $p = (x_p, a_p) \in \mathcal{D}$ carries a set of $m$ numeric attributes
$a_p = \{a_p^1, a_p^2, \dots, a_p^m\}$ alongside its $d$-dimensional vector
$x_p$, and a query bounds a subset of these attributes by intervals. The exact
RFNNS problem is defined as follows.

\begin{definition}[RFNNS]
\label{def:rfnns}
Given a dataset $\mathcal{D}$ with $n$ points $\{p_1, \dots, p_n\}$, a distance
metric $\lVert \cdot, \cdot \rVert$, and a query $(q_i, f_i)$,
where $q_i \in \mathbb{R}^d$ is the query vector and
$f_i = \{\, [l_j, u_j] \mid j \in M \,\}$ denotes the range predicates over a
subset of attributes $M \subseteq \{1, \dots, m\}$, the object of RFNNS is to
identify a $k$-element subset $K_f \subseteq \mathcal{D}$ such that: (1) every
point $p = (x_p, \{a_p^1, \dots, a_p^m\}) \in K_f$ satisfies
$l_j \le a_p^j \le u_j$ for all $j \in M$; and (2) for any point
$u \in \mathcal{D}_f \setminus K_f$,
$\lVert x_p, q_i \rVert \le \lVert x_u, q_i \rVert$.
\end{definition}
Condition~(1) instantiates the matching function of Definition~\ref{def:fknn}
as
\begin{equation}
g(a_p, f_i) \;=\; \prod_{j \in M} \mathbf{1}\!\left[\, l_j \le a_p^j \le u_j \,\right],
\label{eq:range-g}
\end{equation}
so $\mathcal{D}_f$ contains the points satisfying all intervals in $f_i$
simultaneously, and condition~(2) is \eqref{eq:fknn} restricted to that set. A
query with $|M| = 1$ is \emph{single-attribute} and $|M| \ge 2$
\emph{multi-attribute}; we address both in our paper.

\smallskip
\noindent\textbf{Graph construction.}
Graph-based methods are the dominant approach to ANNS on high-dimensional
data~\cite{b10,b11,b12,b24}. We build a Vamana proximity graph~\cite{b11}, which inserts points
incrementally. A beam search collects candidates for each new point, and a pruning step keeps at most $R$ diverse out-neighbors using the $\alpha$-dominance rule of DiskANN. Vector distance alone is a poor basis for
range filtering, since under a narrow interval most out-neighbors of a vertex
are ineligible and the traversal wastes its budget on points that cannot be returned. We therefore adopt the threshold pruning of JAG~\cite{b18}, the prune runs once per threshold $t$ over a capped attribute distance $\max(\mathrm{dist}_A - t,\, 0)$, each pass receiving a degree budget of $R/|T|$.
Thresholds are quantiles of each point's sampled attribute-distance
distribution. A large threshold zeroes every candidate's attribute distance,
reducing the pass to standard Vamana pruning on vector distance; a small one
zeroes only the attribute-nearest candidates, retaining edges among points
likely to co-occur within a narrow range. Since each pass is capped at $R/|T|$
neighbors, the merged out-neighborhood stays bounded by $R$ and the index
occupies $O(nR)$ space, independent of $|T|$, a single graph adapts to
selectivity through its edge mix rather than through duplicated indexes. JAG~\cite{b18} gives a range filter distance for one numeric attribute, but does
not say how several such attributes combine. In our system $\varphi$ is the
count of predicates a candidate violates, and the filter penalty
$\lambda\varphi$ is added to the vector distance to form the score the search
orders by during traversal. Counting weighs every attribute equally and keeps $\varphi$ at most
$m$, so a single $\lambda$ suffices.

\smallskip
\noindent\textbf{Beam search.}
Queries are answered by beam search~\cite{b11,b12}, which maintains a
worklist $W$ of the $L$ closest candidates encountered so far, where $L$
is the \emph{beamwidth}. Each iteration selects the closest unvisited
point in $W$, computes the distances from $x_q$ to its out-neighbors,
inserts them into $W$, and discards the farthest points when $|W|$ exceeds
$L$. The traversal terminates once every point in $W$ has been visited,
and the closest $k$ points in $W$ are returned. The beamwidth bounds both
the cost and the accuracy of the search: each iteration costs at most $R$
distance computations, while a small $L$ discards candidates early and
cannot recover from a descent into the wrong region of $G$.

\subsection{Motivation: Beamwidth vs. Recall}

The beamwidth $L$ bounds the size of the candidate set maintained during
traversal. Each iteration expands the closest unvisited candidate and inserts
its out-neighbors; when the set exceeds $L$ entries, the farthest are
discarded.

For a query $q$, let $K_f$ be the exact filtered $k$ nearest neighbors and
$\tilde{K}_f$ the set returned. Recall@$k$ is $|\tilde{K}_f \cap K_f| / k$,
averaged over the query set. The number of filter attributes $m$ is the number of attributes the query constrains, and a point is valid only if it satisfies all $m$ predicates. As $m$ grows the valid set shrinks and the surviving points lie further apart in the graph, so at a fixed $L$ recall falls with $m$.

A large beamwidth gives high recall and low throughput; a small one gives low recall and high throughput. Both follow from the size of the candidate set the search maintains. $L$ is the capacity of that set. A larger $L$ admits more candidates, and every admitted candidate is eventually expanded, its out-neighbors scored against the query. The number of distance computations per query therefore grows with $L$, and throughput falls in proportion. A candidate dropped from the set when it overflows is never expanded, and if the path to the true neighbors ran through it, that path is closed the traversal converges in a region that does not contain the answer. A large candidate set keeps such nodes available and the search reaches the neighbors it was meant to find; a small one discards them early and returns
whatever it converged on.

\begin{figure}[htbp]
\vspace{-0.085in}
  \centering
  \includegraphics[width=\columnwidth]{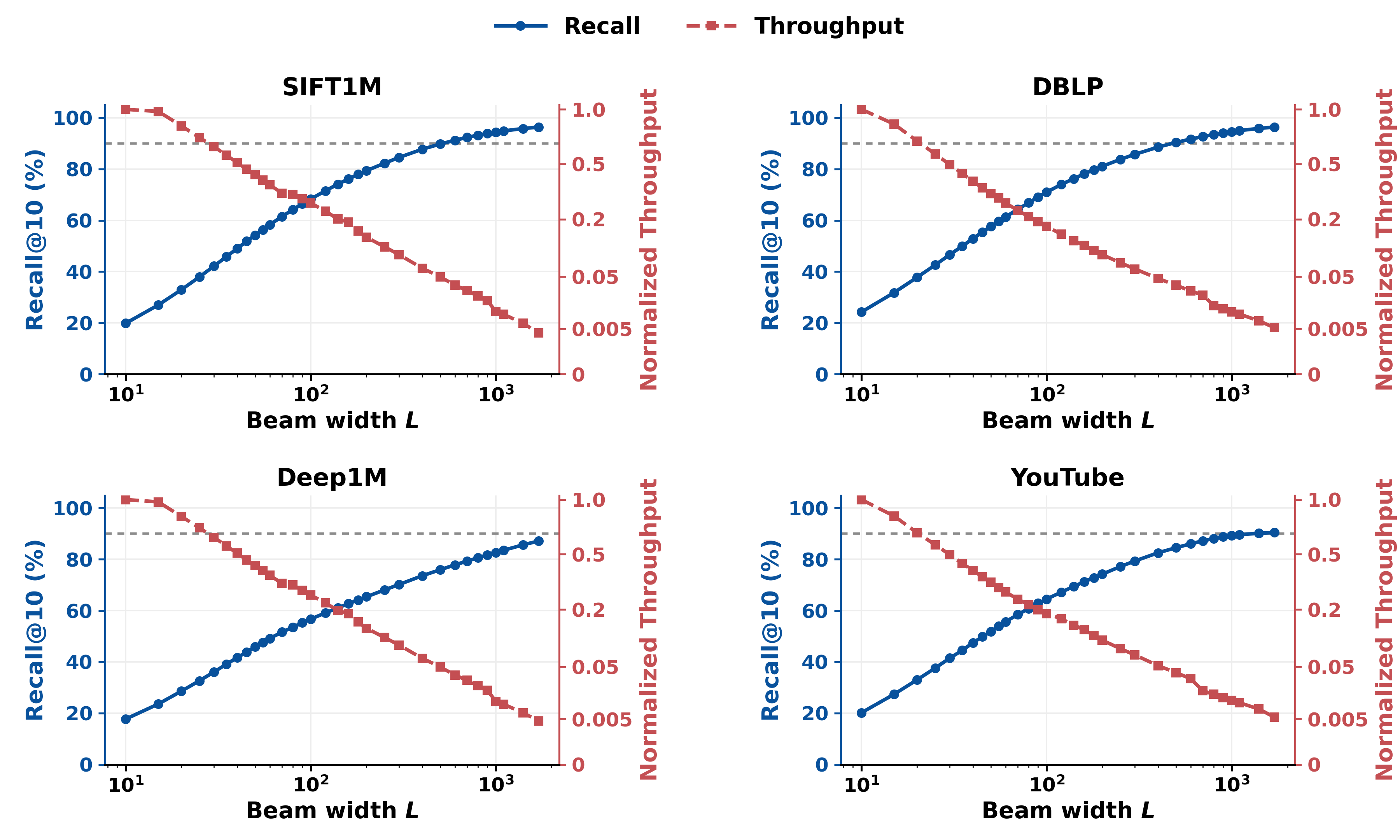}
  \caption{Recall and throughput trade-off with beamwidth $L$ for Garfield~\cite{b17} with $m{=}4$). Throughput is normalized to its maximum over the range of $L$ shown.}
  \label{fig:tradeoff}
\vspace{-0.075in}
\end{figure}

Fig.~\ref{fig:tradeoff} shows this for Garfield~\cite{b17} with $m = 4$, where $m$ is the number of filtered attributes, on SIFT1M, DBLP, Deep1M, and Youtube datasets. Throughput is normalized with respect to the maximum over the range of $L$. The two curves move in opposite directions on every dataset. For example, on Deep1M the search attains $0.21$ recall at $L = 10$, where throughput is at its maximum; reaching $0.9$ recall requires $L = 500$, at a $17\times$ loss in throughput, and reaching $0.97$ requires $L = 1700$, by which point throughput has fallen by more than
$190\times$. The same trade-off appears at $m = 1$ and $m = 2$ .

BANG~\cite{b16} performs unfiltered ANNS on GPU, attaining low recall on filtered workloads regardless of $L$, yet shows the same throughput-recall tradeoff with beamwidth. JAG~\cite{b18} supports only single-attribute filtering and runs on a CPU, it does not support GPU. However, the severity of recall-throughput arises when multi-attribute filtering is supported.

\section{BOA: BeamWidth Online Adaptation}
%

\begin{figure}[!h]
\vspace{-0.085in}
    \centering
    \includegraphics[width=0.9\columnwidth]{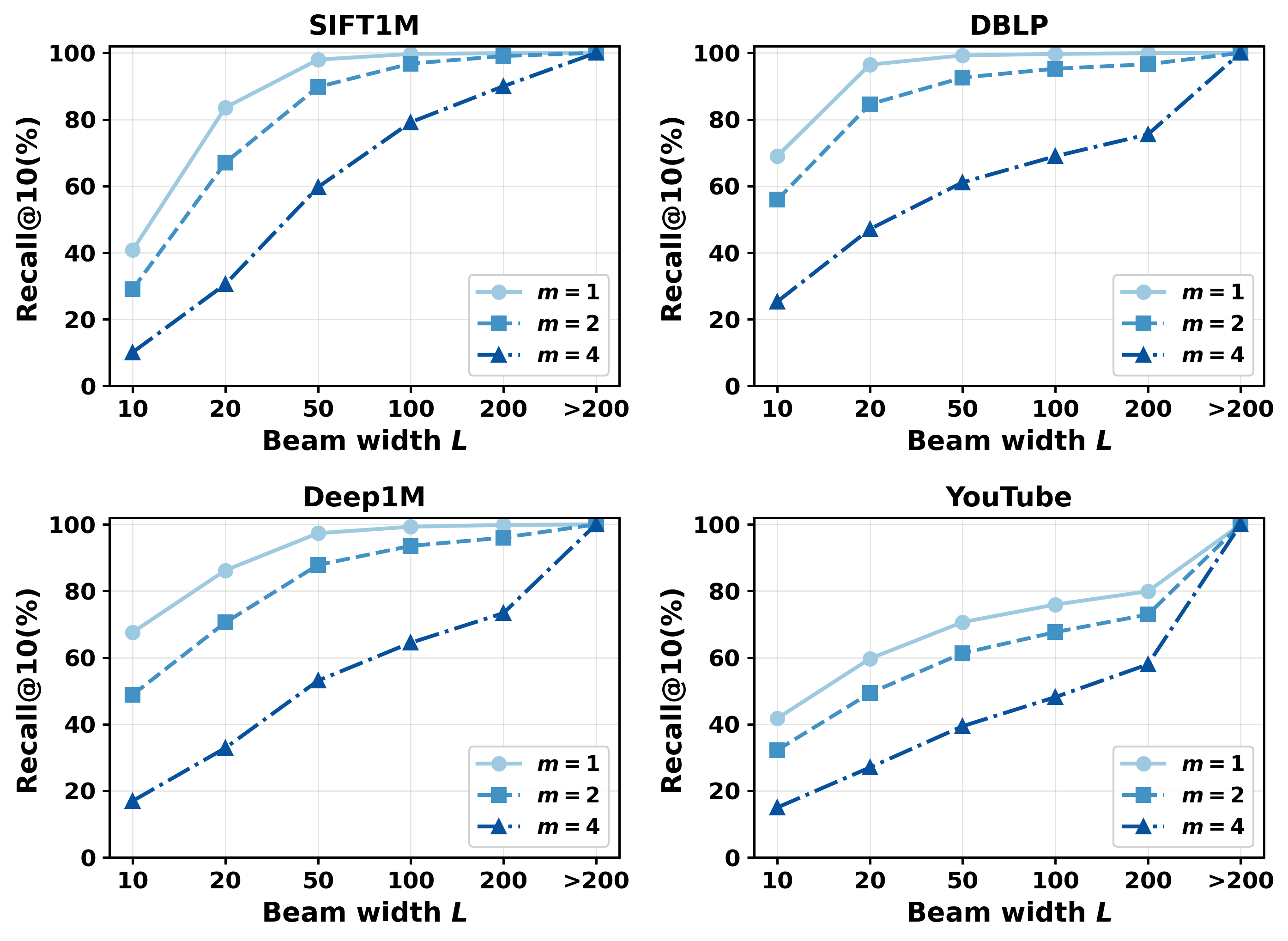}
    \caption{Recall Percentage vs. Bean Width.}
    \label{fig:lneeded}
\vspace{-0.075in}
\end{figure}

We carried a study of multiple datasets that reveal the cause of recall-throughput tradeoff observed in fixed beamwidth systems. Fig.~\ref{fig:lneeded} shows the recall achieved for a large batch of queries across beamwidths ranging from 10 to 200. For a given $L$, recall essentially gives us the percentage of queries that were accurately solved with beamwidth of $L$. From the data in Fig.~\ref{fig:lneeded} we observe that not every query needs a wide beam. In fact in most cases, half of the queries require a beamwidth of 50 or lower. The distribution differ across datasets as they depend upon multiple factors including the dataset, the query workload, where each query's neighbors lie in the graph relative to its filter.

If a query is already solved at $L = 10$, searching it again at $L = 100$ is a \emph{wasteful computation} that adds to the cost but does not improve the recall. Therefore, to obtain high recall across a batch of queries, fixed beamwidth systems pay the high cost on every query in the batch. The wasted computation is not minimal. Widening the beam from $10$ to $500$ costs the throughput to drop by large factors indicating that resources taken by queries whose answers have already stopped changing is substantial.

The above observation motivates our approach based on online \emph{beamwidth adaption} to match the queries. We evaluate a batch of queries over \emph{multiple phases}: search first with a small beamwidth, set aside the queries that are already answered, and advance the rest to search with a wider beamwidth. A query answered in an early phase leaves the batch and requires
no further work, so each phase runs on a strictly smaller query set than the preceding phase. The wide-beam search that in fixed-beamwidth system applies to all $\rho$ queries is applied to only few queries that need it, and the throughput cost of
that width is paid only by a small subset of queries.

The potential benefit can be examined by further analyzing the data in Fig.~\ref{fig:lneeded}. Note that the potential for throughput improvement by adapting $L$ is substantial for all data sets and $m$ values. The higher the value of $m$ the greater is the effort (i.e., $L$ value) required to solve the query. The YouTube data set with $m=4$ has the highest degree of wasted computation and hence the drop in throughput with increasing $L$. At $L = 50$ only 40\% of queries are resolved and to obtain recall greater than 90\%, $L$ should be well over 200. As our experiments presented later show, we can achieve a 94.05\% recall simply by using the average $L$ of 77.

Next we present our beamwidth adaptation-based multi-phase algorithm, multi-attribute filtering algorithm, some optimizations, and finally GPU implementation details. BOA supports multi-attribute filtering across multiple
phases on the GPU, changing both the traversal and the surviving
query set from phase to phase; no prior system combines the two.

\begin{algorithm*}[!t]
\caption{BOA: Beamwidth Online Adaptive Search.}
\label{alg:adaptive}
\begin{algorithmic}[1]
\Require graph $G$; batch of $\rho$ queries $Q$ with filter
  predicates; neighbors $k$; initial beamwidth $L_{\min}$; growth
  factor $\alpha > 1$; confidence threshold $\varepsilon$; residual
  fraction $\tau$; filter penalty $\lambda$
\Ensure $\{K_i\}$: the filter-valid $k$-NN of each $(q_i, f_i) \in Q$
\State $A \gets \{1, \ldots, \rho\}$
  \Comment{active (unresolved) queries}
\State $L \gets L_{\min}$
  \Comment{initial beamwidth}
\While{$|A| > \tau \cdot \rho$}
  \Comment{each iteration is one phase}
  \For{$j \gets 1$ \textbf{to} $|A|$ \textbf{in parallel}}
    \State $Q_A[j] \gets Q[A[j]]$
      \Comment{compact active queries}
  \EndFor
  \State $\{K'_j, D'_j\} \gets
    \textsc{FilterAwareBeamSearch}(G, Q_A, k, L, \lambda)$
  \For{$j \gets 1$ \textbf{to} $|A|$ \textbf{in parallel}}
    \State $K_{A[j]} \gets K'_j$;\;\; $D_{A[j]} \gets D'_j$
      \Comment{scatter results}
  \EndFor
  \ForAll{$i \in A$ \textbf{in parallel}}
    \State $\mathit{confident}_i \gets
      \big(\, \gamma(D_i) \le \varepsilon \,\big)$
      \Comment{confidence test, Eq.~\ref{eq:confidence}}
  \EndFor
  \State $F \gets \{\, i \in A : \mathit{confident}_i \,\}$
    \Comment{finished queries}
  \State $A \gets A \setminus F$
    \Comment{survivors escalate}
  \State $L \gets \alpha \cdot L$
\EndWhile
\State \Return $\{K_i\}$
  \Comment{all queries return results}
\end{algorithmic}
\end{algorithm*}

\begin{algorithm*}[!h]
\caption{Filter-Aware Beam Search.}
\label{alg:fabs}
\vspace{2pt}
\begin{algorithmic}[1]
\Require graph $G$; entry medoid $s$; PQ distance tables; batch of $\rho$
queries $Q$ st each query is of the form $(q_i, f_i)$ where $q_i$ is the query vector and $f_i$ is the filter predicate; neighbors $k$;
beamwidth $t\,(\geq k)$; filter penalty $\lambda$
\Ensure $\{K_i\}$: the filter $f_i$-valid $k$ nearest neighbors of each query $(q_i, f_i) \in Q$
\ForAll{$(q_i, f_i) \in Q$ \textbf{in parallel}} \Comment{one thread-block per query}
    \State $L_i \gets \{s\}$; \; $u_i \gets s$; \; $converged \gets \textbf{false}$
    \While{\textbf{not} $converged$}
        \State $N_i \gets \textsc{FetchNeighbors}(u_i, G)$ \Comment{graph stays on CPU, only current node's neighbor IDs are sent to the GPU}
        \ForAll{$n \in N_i$}
            \State $\varphi_i[n] \gets \mathrm{dist}_F(a_n, f_i)$ \Comment{filter distance}
        \EndFor
        \State transfer $N_i$ and $\varphi_i$ to GPU
        \State $N'_i \gets \{\, n \in N_i : \textbf{not }\textsc{Visited}(i,n) \,\}$; \; mark $N'_i$ visited \Comment{Bloom filter on GPU, candidates seen before dropped}
        \ForAll{$n \in N'_i$ \textbf{in parallel}} \Comment{on GPU}
            \State $D_i[n] \gets \textsc{PQDist}(n, q_i) + \lambda \cdot \varphi_i[n]$ \Comment{filter-penalized distance}
        \EndFor
        \State $(\hat{D}_i, \hat{N}_i) \gets \textsc{ParallelSort}(D_i, N'_i)$ \Comment{on GPU}
        \State $L_i \gets \textsc{ParallelMerge}(L_i, \hat{N}_i, \hat{D}_i)$ \Comment{on GPU}
        \State \textbf{if} $|L_i| > t$ \textbf{then} keep the $t$ closest entries of $L_i$
        \State $u_i \gets$ nearest unvisited node in $L_i$; \; mark $u_i$ visited
        \State $converged \gets (\text{all nodes in } L_i \text{ are visited})$
    \EndWhile
    \State $K_i \gets$ the $k$ filter-valid ($\varphi = 0$) nodes in $L_i$ nearest to $q_i$
\EndFor
\State \Return $\{K_i\}$
\end{algorithmic}
\end{algorithm*}

\subsection{Adaptive Filtered Search Algorithm}
\underline{\textsf{How adaptation works.}} Our adaptive algorithm, presented in Algorithm~\ref{alg:adaptive}, iteratively evaluates a batch of $\rho$ queries, where each iteration is a phase. As the algorithm progresses through phases, the beamwidth used in the search is increased. Moreover, in each phase, only a subset of queries from the preceding phase that remain unresolved is evaluated. That is, fewer and fewer queries are evaluated at increasingly higher beamwidths as the algorithm progresses through the phases.

In the first phase, all queries are active and they are evaluated with an initial beamwidth of $L_{\min}$ (lines 1--2). A phase begins by gathering the active queries into consecutive positions (lines 4--5). Algorithm~\ref{alg:fabs} searches for gathered queries at current width (line 6) and writes back the results (lines 7--8). Note that the results of queries are maintained according to the index associated with them during the first phase regardless of the phase that produces the results.

After active queries have been evaluated in a given phase, we separate them into those that have now been resolved (line 11) and remaining survivors (line 12) that will remain active for evaluation in the next phase. The survivors will then be evaluated in the next phase at a beamwidth that is a factor of $\alpha$ greater than the beamwidth of the just concluded phase (line 13). 
Separating resolved queries from surviving ones is carried out using the confidence test (line 10), where the sorted candidate list phase produced before its nearest $k$ entries are taken as the result. That list holds more filter-valid candidates than the $k$ returned, and a query is
\emph{confident} when the relative gap between the $k$-th and the farthest of them is small:
{\setlength{\abovedisplayskip}{3pt}\setlength{\belowdisplayskip}{3pt}
\setlength{\abovedisplayshortskip}{0pt}\setlength{\belowdisplayshortskip}{0pt}
\begin{equation}
  \gamma(D_i) = \frac{d_i^{\max} - d_i^{(k)}}{d_i^{\max}} \le \varepsilon .
  \label{eq:confidence}
\end{equation}}
The gap measures how far the search reaches beyond the $k$ it will return, and is normalized by $d_i^{\max}$ because absolute distances differ in scale across datasets. A narrow gap means the remaining candidates sit at essentially the
distance of the $k$th, so a wider beam would only add points farther than those already returned and the top-$k$ is settled. A wide gap means the search has not converged on a single neighborhood, so a wider beam may reach points
nearer than the current $k$-th and the query escalates. We use $\varepsilon = 0.05$ for every data set.

\emph{Note that the number of phases is not fixed. How many phases a given query takes is determined at run time by its difficulty. Most of the queries maybe answered at a narrow beam, while the remaining whose neighborhoods lie in sparse or in heavily filtered regions take multiple phases. Moreover, the cost widest beams are paid for by fewest queries since each phase runs only on survivors from preceding phase.}

The loop terminates when fewer than $\tau\rho$ queries remain active. A small residual fraction likely does not converge at any beamwidth the system is run on, and $\tau$ bounds the cost of pursuing it; those queries return the best answer found. Every query returns a result (line 14).

\underline{\textsf{How Filtered Search Works.}} Next we discuss the details of the filter-aware beam search presented in Algorithm~\ref{alg:fabs}. This algorithm evaluates a batch of $\rho$ queries at a single beamwidth $t$, and is invoked by Algorithm~\ref{alg:adaptive} once per phase.

Since the queries being solved are independent, each query is assigned its own thread block and the whole batch traverses the graph concurrently (line 1). Every query enters at the same medoid $s$ and maintains its own worklist $L_i$ of the $t$ closest candidates found so far (line 2).

Each iteration expands one node. The adjacency list of the current candidate is read on the CPU, where the graph resides (line 4), and in the same pass over that memory the filter distance $\varphi_i[n]$ is evaluated for every neighbor: the number of the query's range predicates that neighbor violates, so $\varphi = 0$ exactly when it is filter-valid (lines 5--6). Identifiers and filter distances are transferred to the GPU
together in a single asynchronous copy, which keeps each candidate paired with its own violation count (line 7). A per-query Bloom filter then discards candidates seen in an earlier iteration (line 8); without it, repeated entries would displace better candidates from the worklist and end the traversal prematurely.

The surviving candidates are \emph{scored} on the GPU against the product-quantized codes, and the filter enters the score directly (lines 9--10): a candidate violating $v$ of the query's predicates is ranked at $\textsc{PQDist}(n, q_i) + \lambda v$. Because $\lambda$ is set well above the scale of vector distances, valid candidates are ordered ahead of invalid ones, and vector distance breaks ties within each group. Consequently, invalid candidates are demoted rather than removed. Thus, they remain in the worklist and can still be expanded, allowing traversal to pass through invalid regions to reach valid ones. Removing them would have disconnect the graph precisely where the predicate is most selective, since under a narrow interval the valid points are sparse and frequently reachable only via invalid intermediaries.

The scored candidates are sorted and merged into the worklist (lines 11--13). Both operations run in parallel across the thread block. In the merge, each element of the two sorted lists is assigned a thread, which binary-searches its element's position in the opposite list; the sum of that position and the element's own index gives its slot in the merged output, so every thread writes directly to its destination. The sort for the new candidate list applies the same merge routine bottom-up, beginning from singleton lists and doubling the merged length each round, so it completes in $\log |N'_i|$ rounds. Both keep their lists in shared memory, which is possible because a candidate list holds at most $R+1$ entries.

The worklist is then truncated to its $t$ closest entries (line 13), which bounds the work each query performs. The next node to expand is the nearest unvisited entry of the worklist (line 14), and the traversal ends for a query once every entry has been expanded (line 15). The predicate is enforced exactly at the end: only candidates with $\varphi = 0$ are eligible for the result, and the $k$ nearest of those are returned (line
16). The penalty guides navigation; it does not decide validity.

\smallskip
\underline{\textsf{Optimizations.}} Next, we describe a couple of optimizations that enable us to maximize the throughput: \emph{redundancy avoidance}; and \emph{pipelining batches}.

\paragraph*{Redundancy Avoidance} Consider a query that is evaluated over two consecutive phases with $L$ and $\alpha L$ beamwidths. We should avoid repeating the computation performed with beamwidth $L$ during the next phase that uses $\alpha L$. This is achieved by computing the $topk_L$ results from the first $L$ children of each node during first phase graph traversal. In the next phase we compute the $topk_{\alpha L - L}$ results from next $\alpha L - L$ children. Finally, by comparing $topk_L$ and $topk_{\alpha L - L}$, we obtain $topk_{\alpha L}$ as the results for all $\alpha L$ children.

\begin{figure}[t]
  \centering
  \includegraphics[width=1.05\columnwidth]{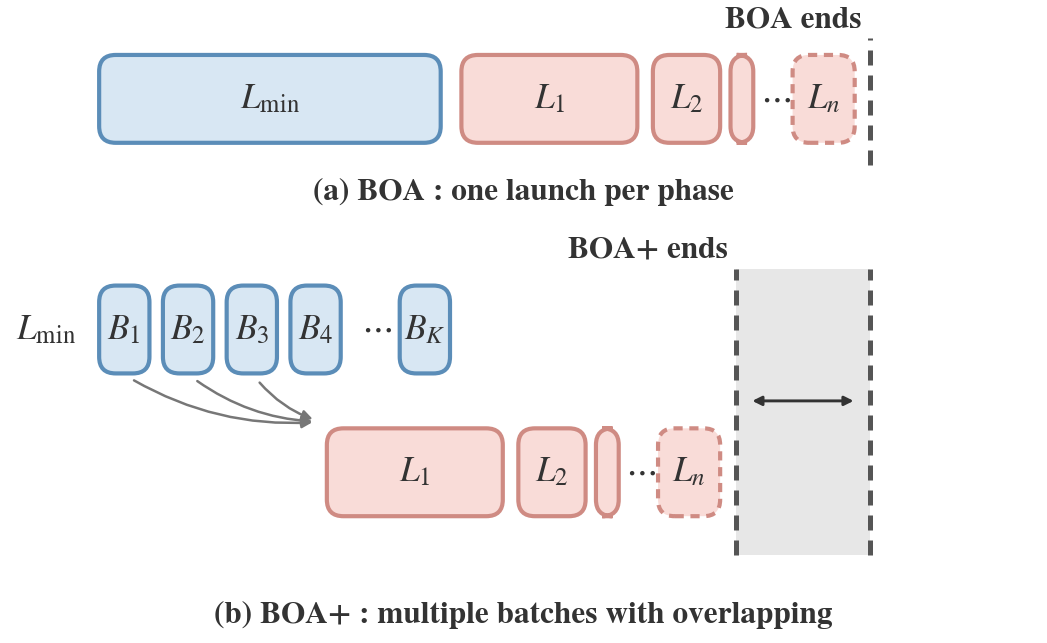}
  \caption{\textsf{BOA} vs.\ \textsf{BOA+} execution schedule.}
  \label{fig:overlap}
  \vspace{-0.175in}
\end{figure}

\paragraph*{Pipelining Multiple Batches}
Because the search proceeds in phases, part of one phase can overlap
another. In BOA it does not, as Fig.~\ref{fig:overlap}(a) shows: a phase
cannot begin until the preceding one ends, since the confidence test
needs the distances that phase produced, and the $L_{\min}$ pass covers
all $\rho$ queries in a single launch. The dependency, however, is
between phases of the same query, not between different queries.
\textsf{BOA+} splits only the first. It issues the $L_{\min}$ search as $K$
sub-batches $B_1 \ldots B_K$, so the survivors of the early sub-batches
are known before the later ones have run. Those survivors are
accumulated and escalated together at the wider beamwidth in one pooled
launch, concurrently with the $L_{\min}$ search of the sub-batches that
remain Fig.~\ref{fig:overlap}(b). Escalating each sub-batch as it
finishes would split a small query set across many launches, none large
enough to fill the GPU. Pooling them keeps the escalation wide. Both
settle the same batch to the same recall. \textsf{BOA+} only reaches that point
sooner, and the shorter span is where the higher throughput comes from.

\subsection{\textsf{BOA} Implementation on a GPU System}


A graph traversal is a sequence of dependent steps: the neighbors expanded in one iteration determine which node is expanded in the next. Batching thousands of independent queries recovers the parallelism a single traversal lacks, but it does not remove the dependency within a query, and at scale each step requires data the GPU does not hold. Designing the execution so that neither processor waits on the other is therefore the central concern.

\begin{figure}[htbp]
\centering
\begin{tikzpicture}[
  font=\scriptsize,
  cblk/.style={draw=black!40,rounded corners=2pt,fill=cpublk,
               minimum width=2.3cm,minimum height=0.5cm,align=center,inner sep=2pt},
  gblk/.style={draw=black!40,rounded corners=2pt,fill=gpublk,
               minimum width=2.3cm,minimum height=0.5cm,align=center,inner sep=2pt},
  mblk/.style={draw=black!40,rounded corners=2pt,fill=memblk,
               minimum width=1.6cm,minimum height=0.5cm,align=center,inner sep=2pt},
  num/.style={circle,fill=black!15,inner sep=0.6pt,font=\tiny,minimum size=3mm},
  ar/.style={-{Latex[length=1.2mm,width=1mm]},black!50,thin}
]

\node[cblk] (c1) {Gather neighbors};
\node[cblk,right=0.35cm of c1] (c2) {Filter distance};
\node[num] at (c1.north west) {1};
\node[num] at (c2.north west) {2};
\node[mblk,below=0.4cm of c1] (m1) {Graph index};
\node[mblk,right=0.35cm of m1] (m2) {Original vectors};
\node[draw=black!30,rounded corners=3pt,fit=(c1)(c2)(m1)(m2),inner sep=4pt] (cbox) {};

\node[gblk,below=1.3cm of m1.south,anchor=north] (g1) {Suppress visited};
\node[gblk,right=0.35cm of g1] (g2) {Penalized PQ dist.};
\node[gblk,below=0.35cm of g1] (g3) {Sort and merge};
\node[gblk,right=0.35cm of g3] (g4) {Re-rank exact};
\foreach \i/\n in {g1/3,g2/4,g3/5,g4/6}
  \node[num] at (\i.north west) {\n};
\node[mblk,below=0.35cm of g3] (r1) {PQ codes};
\node[mblk,right=0.35cm of r1] (r2) {Attribute array};
\node[draw=black!30,rounded corners=3pt,fit=(g1)(g2)(r1)(r2),inner sep=4pt] (gbox) {};

\draw[ar] (c1) -- (c2);
\draw[ar] (m1) -- (c1);
\draw[ar] (g1) -- (g2);
\draw[ar] (g2.south) -- ++(0,-0.15) -| (g3.north);
\draw[ar] (g3) -- (g4);
\draw[{Latex[length=1.2mm]}-{Latex[length=1.2mm]},black!50,thin] (r1.north) -- (g3.south);

\draw[ar] ($(cbox.south)+(-0.9,0)$) -- node[left,font=\tiny]{IDs, $\varphi$}
          ($(gbox.north)+(-0.9,0)$);
\draw[ar] ($(gbox.north)+(0.9,0)$) -- node[right,font=\tiny]{candidate}
          ($(cbox.south)+(0.9,0)$);

\begin{scope}[on background layer]
  \node[draw=black!40,rounded corners=4pt,fill=cpubg,fit=(cbox),
        inner sep=5pt,label={[yshift=-2pt,font=\scriptsize]above:CPU}] {};
  \node[draw=black!40,rounded corners=4pt,fill=gpubg,fit=(gbox),
        inner sep=5pt,label={[yshift=-2pt,font=\scriptsize]above:GPU}] {};
\end{scope}
\end{tikzpicture}
\caption{\textsf{BOA} workflow across CPU and GPU.}
\label{fig:pipeline}
\vspace{-0.15in}
\end{figure}
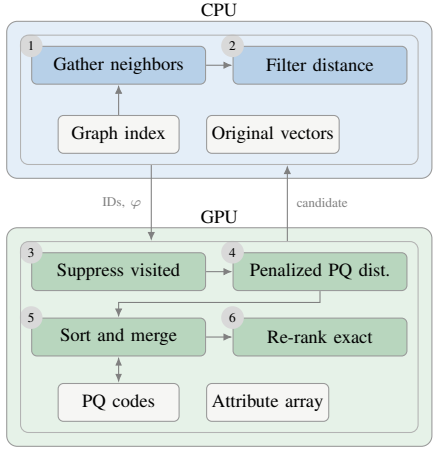

The CPU memory retains the graph index and the original vectors,
laid out in interleaved records so that a node's vector sits immediately
before its neighbor list and one sequential read retrieves both. The GPU
memory retains the product-quantized codes~\cite{b19}, the quantization
pivots, and the per-point attribute array. PQ distance evaluation, candidate
ranking, worklist maintenance, and the final exact-distance pass run on the
GPU, where as adjacency lookup and filter evaluation run on the CPU, where the data they read already resides. Transfers are issued asynchronously on dedicated
streams and overlapped with kernel execution, so that per-iteration latency is
absorbed into work the GPU performs regardless.

As shown in Fig.~\ref{fig:pipeline}, a batch of queries proceeds as follows.
The CPU \circled{1} transfers the query vectors and their range bounds, after
which the GPU \circled{2} builds the PQ distance table~\cite{b19}, giving each
query the distance from its subvectors to every centroid in every subspace.
The traversal then begins. In each iteration the CPU \circled{3} reads the
adjacency list of the current candidate for every active query and, in the
same pass over that memory, evaluates how many of the query's range predicates
each neighbor violates. Identifiers and violation counts are \circled{4}
transferred together in one asynchronous copy. The GPU \circled{5} discards
neighbors already visited and compacts the remainder, \circled{6} computes
their penalized distances as the asymmetric PQ distance~\cite{b19} the sum
of table lookups indexed by the neighbor's quantized code as well as  the
transferred violation count, and \circled{7} sorts them and merges them into
the worklist, selecting the next candidate. On convergence the GPU
\circled{8} ranks the accumulated candidates by exact distance under a hard
predicate penalty, and the top-$k$ identifiers are \circled{9} returned to the
CPU.

Queries within a batch of $\rho$ queries are independent, so  $\rho$ blocks are launched with each query occupying one block; throughput scales with $\rho$ until the hardware saturates. Within a block the width of the work varies across stages \circled{5}--\circled{7}: visited-set
suppression assigns one thread per candidate, PQ distance evaluation assigns a group of eight threads per candidate so that partial sums over the quantization subspaces reduce within a sub-warp, sorting assigns one thread per element of the candidate list, and merging assigns one thread per element of the two lists being combined. Issuing the iteration as a single fused kernel would fix the launch configuration at the widest, leaving threads idle in the rest, so each stage is a separate kernel configured for
its own occupancy.

The phase transition of \textsf{BOA} requires one further
mechanism on the GPU. After the confidence test the survivors are scattered
arbitrarily across the batch index space, and dispatching a thread block per
query so that most terminate on their first instruction would consume
scheduling slots and depress occupancy. We therefore gather the survivors'
query vectors, range bounds, and search state into contiguous arrays, so that
the second phase launches one block per survivor and its cost tracks the
number of queries that require it rather than the size of the batch.

\section{Experiments}
To demonstrate the benefits of our approach we present results for the following three versions of GPU algorithms: 
\begin{itemize}
    \item \textbf{\textsf{Fixed L}} is the baseline that uses a high beamwidth of 500 to achieve high recall of $>$90\%; 
    \item \textbf{\textsf{BOA}} performs online adaptation of beamwidth (L). The multiple-phases with beamwidths of 10, 100, and 200 are used to achieve high recall of $>$90\%; and
    \item \textbf{\textsf{BOA+}} is the version of BOA that is enhanced by incorporating pipelining.
\end{itemize} 
All experiments are performed on a system equipped with 2$\times$ AMD EPYC 7543 32-Core CPUs (64 cores) and an NVIDIA A100 GPU (80GB). BOA is implemented in C++/CUDA using CUDA 12.4. We evaluate the algorithms on four datasets widely used in related work~\cite{b25,b26,b27,b23}; SIFT1M~\cite{b34}, Deep1M~\cite{b37}, DBLP~\cite{b36}, and YouTube~\cite{b35}. Because SIFT1M and Deep1M include vectors but no attributes, synthesize a randomly generated integer attribute for each vector, following standard practice~\cite{b29,b27,b23}. DBLP and YouTube, in contrast, pair real-world vectors with naturally skewed numeric attributes. Following Garfield's approach~\cite{b17}, each query poses a range predicate per attribute, with selectivity generated uniformly between 1\% and 100\%. Table~\ref{tab:datasets} provides a statistical summary of all datasets.
\begin{table}[htbp]
\centering
\caption{Details of Datasets.}
\label{tab:datasets}
\begin{tabular}{lrrrl}
\toprule
\!\!\textbf{Dataset}\!\!&\!\!\textbf{Dim.}\!\!&\!\!\!\!\textbf{\#Base}\!\!&\!\!\textbf{\#Queries}\!\!&\!\!\textbf{Attributes}\!\! \\
\midrule
Deep1M     & 96   & 1,000,000   & 10,000 & Uniform random \\
SIFT1M     & 128  & 1,000,000   & 10,000 & Uniform random \\
DBLP       & 768  & 1,000,000   & 10,000 & Year,\:\#authors,\:\#refs,\:\#cites\!\!\\
YouTube    & 1024 & 1,000,000   & 10,000 & Year, time, \#views, \#likes \\
\bottomrule
\end{tabular}
\vspace{-0.15in}
\end{table}

\subsection{BOA: Recall and Throughput}
\begin{figure}[t]
  \centering
  \begin{subfigure}{0.85\linewidth}
    \centering
    \includegraphics[width=\linewidth]{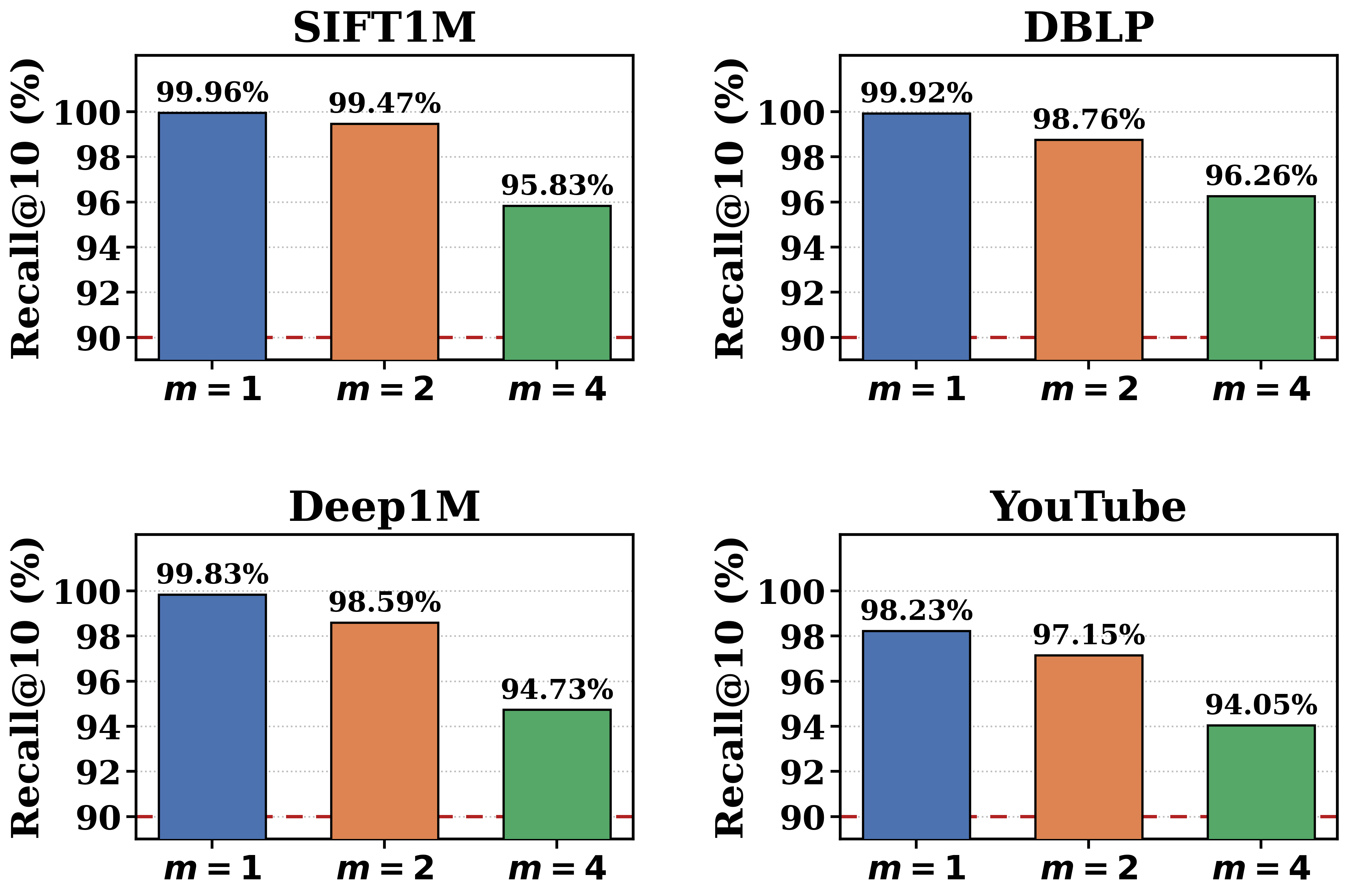}
    \caption{Recall@10}
    \label{fig:exp1_recall}
  \end{subfigure}
  \\[1ex]
  \begin{subfigure}{0.85\linewidth}
    \centering
    \includegraphics[width=\linewidth]{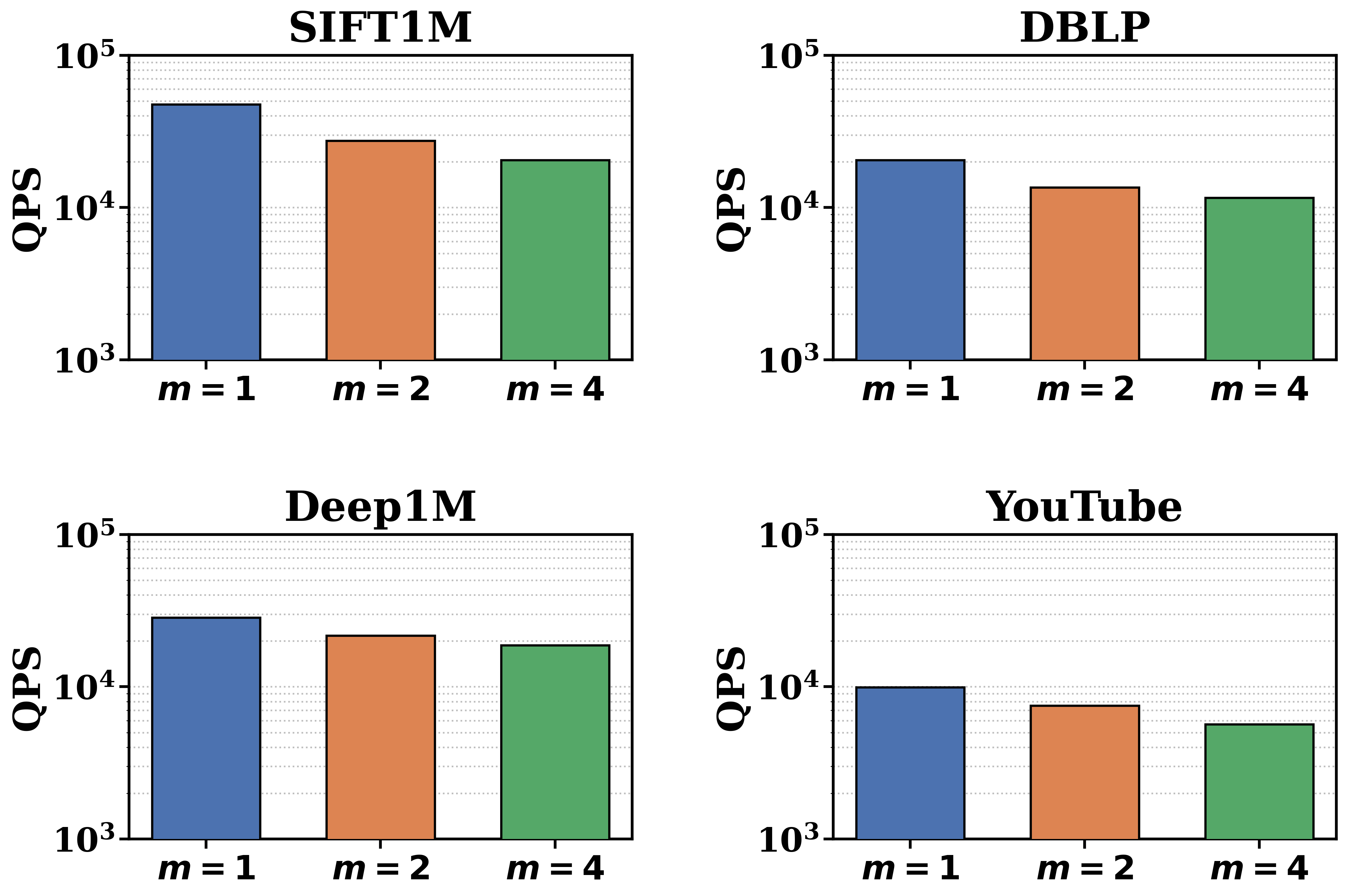}
    \caption{End-to-end throughput}
    \label{fig:exp1_qps}
  \end{subfigure}
  \caption{\textsf{BOA}: Recall@Top10 (\%) and End-to-end Throughput in Queries Per Second (QPS) for one batch of 10{,}000 queries.}
  \label{fig:recall_qps}
  \vspace{-0.15in}
\end{figure}

The performance of \textsf{BOA}, both Recall and Throughput, are plotted in Fig.~\ref{fig:recall_qps} for $m \in \{1, 2, 4\}$ and all four datasets. We evaluate a single batch of 10{,}000 queries per dataset. The beamwidths used are: 10, 100, and 200.

\textit{\textsf{Recall.}}
Fig.~\ref{fig:recall_qps}(a) reports recall@10 for $m = 1, 2, 4$ for all datasets. The pattern is the same everywhere. Recall is highest with a single filter attribute and degrades as multi-attributes are added. At one end is SIFT1M that goes from 99.96\%$\rightarrow$99.47\%$\rightarrow$97.69\% and at the other end is YouTube that goes from 97.15\%$\rightarrow$96.23\%$\rightarrow$94.05\%. 
All twelve values exceed the 90\% target. 

\textit{\textsf{Throughput.}}
Fig.~\ref{fig:recall_qps}(b) reports queries per second 
for $m = 1, 2, 4$. As expected, throughput declines as $m$ increases, since a larger fraction of queries require additional work. At one end is SIFT1M that goes from 47.6k$\rightarrow$27.5k$\rightarrow$20.5k and at the other end is YouTube goes from 13.6k$\rightarrow$7.5k$\rightarrow$5.7k.

Next we present more detailed analysis to demonstrates how our approach consistently and simultaneously delivers high recall and high throughput.

\begin{figure*}[t]
  \centering
  \includegraphics[width=0.715\textwidth]{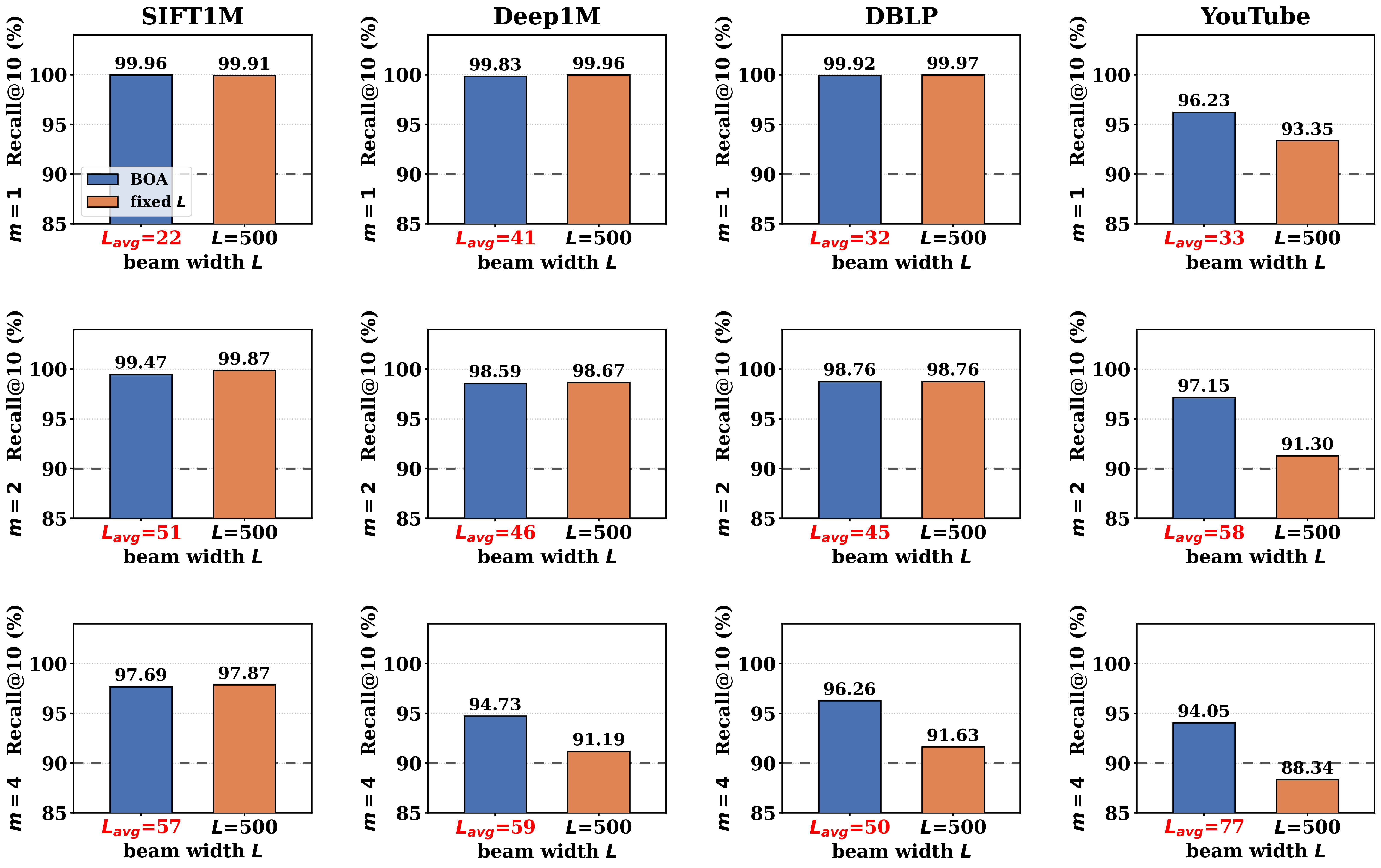}
  \caption{Recall@Top10 at our average beamwidth $L_{avg}$ versus a fixed beamwidth $L=500$.}
  \label{fig:fixed_vs_adaptive}
  \vspace{-0.175in}
\end{figure*}

\subsection{\textsf{BOA}: Achieving Steady High Recall}
Table~\ref{tab:split} shows how many phases are employed for each dataset and different number of attribute filters. From these detailed results we observe the following:
\begin{itemize}\itemsep1.5pt
\item \textit{Number of Phases:}  We observe that three datasets of SIFT1M, Deep1M, and DBLP require two phases with $L$ values of 10 and 100 to achieve high recall rate of greater than 90\%. YouTube, the most challenging workload with 1K vectors, requires three phases with $L$ values of 10, 100, and 200. The recall rate is consistently greater than 90\% and quite a bit higher: 94.05\% to 99.96\%.
\item \textit{Queries Resolved Across Phases:} Depending upon the hardness of the queries, our algorithm automatically adapts the load distribution across the phases. For example, for $m=4$ and SIF1M dataset, \textsf{BOA} resolves 47.9\% and 52.1\% queries in first and second phases respectively. In contrast, for YouTube dataset, \textsf{BOA} resolves 57.0\%, 14.8\%, and 28.2\% of queries in the three phases.
\end{itemize}

\begin{table}[!h]
\centering
\caption{\textsf{BOA} Behavior: Percentage of queries from a 10{,}000-query batch resolved at various beamwidths across up to three phases and the resulting Recall@10.}
\label{tab:split}
\begin{tabular}{llrrrr}
\toprule
Dataset & $m$ & $L{=}10$ & $L{=}100$ & $L{=}200$ & Recall@10 \\
\midrule
\multirow{3}{*}{SIFT1M}  & 1 & 86.1 & 13.9 & --   & 99.96\%\\
                         & 2 & 54.3 & 45.6 & --   & 99.47\%\\
                         & 4 & 47.9 & 52.1 & --   & 97.69\%\\
\midrule
\multirow{3}{*}{Deep1M}  & 1 & 65.7 & 34.2 & --   & 99.83\%\\
                         & 2 & 59.8 & 40.1 & --   & 98.59\%\\
                         & 4 & 45.4 & 54.5 & --   & 94.73\%\\
\midrule
\multirow{3}{*}{DBLP}    & 1 & 75.7 & 24.2 & --   & 99.92\% \\
                         & 2 & 61.4 & 38.5 & --   & 98.76\%\\
                         & 4 & 55.4 & 44.5 & --   & 96.26\%\\
\midrule
\multirow{3}{*}{YouTube} & 1 & 82.0 & 11.6 & 5.2  & 97.15\% \\
                         & 2 & 67.8 & 13.8 & 18.2 & 96.23\%\\
                         & 4 & 57.0 & 14.8 & 28.2 & 94.05\%\\
\bottomrule
\end{tabular}
\vspace{-0.15in}
\end{table}

\vspace{-0.1in}
\subsection{Beamwidth Needed for High Recall: \textsf{BOA} vs. \textsf{Fixed L}}
\vspace{-0.05in}
Table~\ref{tab:split} clearly demonstrated that BOA adapts dynamically to the workload and always delivers high recall. Next we show that this adaptation greatly reduces the average beamwidth ($L_{avg}$) used across all queries. Fig.~\ref{fig:fixed_vs_adaptive} plots the recall rate of BOA and non-adaptive Fixed L baseline. Note that $L_{avg}$ ranges are: 22 to 41 for $m=1$; 45 to 58 for $m=2$; and 50 to 77 for $m=4$. In contrast, Fixed L uses beamwidth of 500 and still delivers similar or lower recall. In other words, BOA's adaptivity is effective in delivering a greatly reduced $L_{avg}$ while delivering similar or superior recall.

\begin{figure}[!t]
\vspace{-0.05in}
  \centering
  \includegraphics[width=0.925\linewidth]{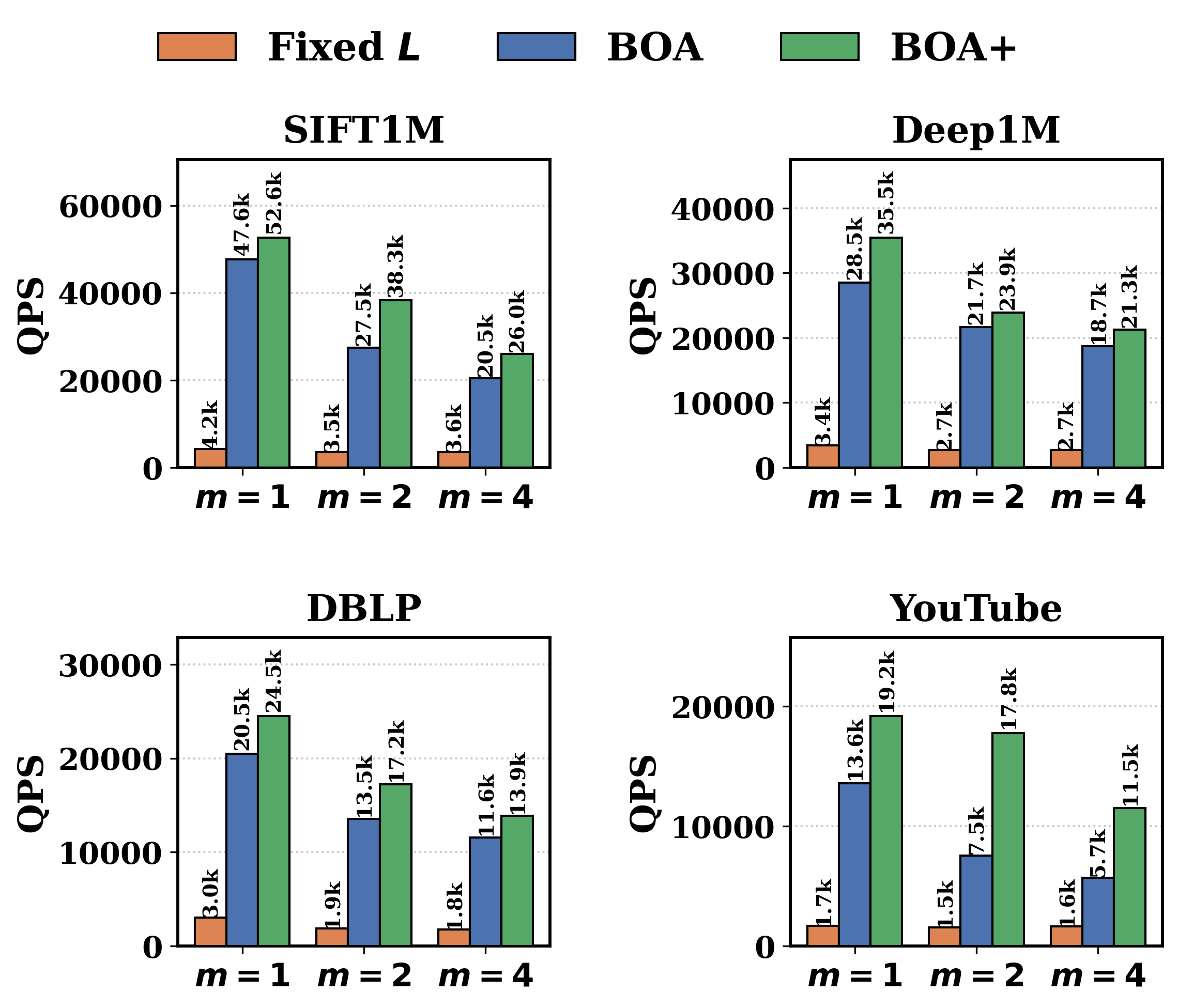}
\caption{Throughput: \textsf{BOA} and \textsf{BOA+} vs. \textsf{Fixed L}.}
\label{fig:pipelining}
\vspace{-0.2in}
\end{figure}

\subsection{Throughput Enhancement: \textsf{BOA} and \textsf{BOA+} vs. \textsf{Fixed L}} 
Finally, we show that low $L_{avg}$ leads to significant improvement in throughput. We compare the throughput of \textsf{Fixed L} with \textsf{BOA} and \textsf{BOA+}. While \textsf{BOA} solves all 10,000 queries in one batch, \textsf{BOA+} divides the queries into 10 batches of 1,000 queries each and exploits pipelining.

Fig.~\ref{fig:pipelining} and Table~\ref{tab:factor} show that throughput of \textsf{BOA} is 3.5$\times$ to 11.3$\times$ times greater than for \textsf{Fixed L}. Moreover, pipelining further improves the throughput significantly resulting in 7.0$\times$ to 12.5$\times$ improvements in throughput over \textsf{Fixed L}.

\begin{table}[!h]
\centering
\caption{Throughput Improvement Factor Over \textsf{Fixed L}.}
\label{tab:factor}
\vspace{-0.1in}
\begin{tabular}{llrrr} \\
\toprule
        & Alg. & \multicolumn{3}{c}{Throughput Improvement} \\
Dataset & Version & $m{=}1$ & $m{=}2$ & $m{=}4$ \\
\midrule
\multirow{2}{*}{SIFT1M}  & BOA+ &12.5$\times$ &10.9 $\times$ & 7.2$\times$ \\
                         & BOA & 11.3$\times$ & 7.8$\times$ & 5.7$\times$\\
\midrule
\multirow{2}{*}{Deep1M}  & BOA+ & 10.4$\times$ &  9.0$\times$ & 7.9$\times$ \\
                         & BOA  &  8.4$\times$ &  8.1$\times$ & 6.9$\times$ \\
\midrule
\multirow{2}{*}{DBLP}    & BOA+ &  8.1$\times$ &  9.3$\times$ & 7.9$\times$ \\
                         & BOA  &  6.7$\times$ &  7.3$\times$ & 6.6$\times$ \\
\midrule
\multirow{2}{*}{YouTube} & BOA+ & 11.4$\times$ & 11.5$\times$ & 7.0$\times$ \\
                         & BOA  &  8.1$\times$ &  4.9$\times$ & 3.5$\times$ \\
\bottomrule
\end{tabular}
\vspace{-0.1in}
\end{table}

\section{Related Work}
\vspace{-0.025in}
There is a great deal of research on vector search that has been carried out on evaluating approximate nearest neighbor queries\cite{b21,b24}. These works evaluate queries in batches to fully exploit hardware parallelism present in CPUs and GPUs and maximize throughput ~\cite{b13,b14,b15,b16,b17,b18, b39,b43}. The use of GPU is attractive because for massive parallelism that when effectively exploited leads to high throughput. The algorithms and systems differ in following respects:

\textsf{Filtering Attributes.} While some systems support multi-attribute filtering~\cite{b17}, others support only a single attribute filter~\cite{b16}, yet others do not support filtering attributes at all. Nearly all of this filtered work targets the CPU. Filtered search on the GPU has seen far less work despite the throughput it offers. In this work we considered range filters. However, there are other types of filters, namely equality and subset filters ~\cite{b8,b37,b39,b40}. Our work can be be extended to consider these filter types in a straightforward manner.

\textsf{Graph construction.} In presence of filtering attributes, there are two approaches that have been proposed. One approach that is used by \textsf{JAG}~\cite{b18} (for CPU) and \textsf{BOA} (for GPU) constructs a single attribute-unaware graph offline, and then handles filtering during the traversal process. Another approach, used by Garfield~\cite{b17}, builds an attribute-aware partitioned graph such that filter attributes associated with a query are used to index the graph partition at the start of the search. Then search is performed mostly within the partition, and to a much lesser extent by traversing inter-partition edges and traversing other graph partitions. SeRF~\cite{b22}, iRangeGraph~\cite{b23}, UNIFY~\cite{b27}, and DIGRA~\cite{b29} follow this route with segment graphs, segment trees, segmented inclusive graphs, multi-way trees, and KD-trees respectively, and all of them inflate the index. GAAF~\cite{b38} instead maintains an ensemble of graphs partitioned by label frequency and routes each query to the subset whose labels it matches. Systems also differ in when the filter is applied, with pre-filtering restricting the search before it begins~\cite{b31}, post-filtering discarding failures after retrieval~\cite{b42}, and both of the above folding the check into traversal. PANNS~\cite{b43} also splits the search into two phases by hardness, entering the second once the top-$k$ entries of a query's beam are all visited and then widening the step size while reducing the cut-off factor. Its adaptation is limited to this switch within a single traversal, and it supports neither filtering nor execution on a GPU.


\section{Conclusion}
The Approximate Nearest Neighbor Search (ANNS) on high-dimensional vectors is a core operation in modern recommendation systems. This problem becomes computationally more expensive when multi-attribute filtering is supported as deeper search is needed for many queries that is achieved by setting beamwidth to a very high number for all queries. This approach yields high recall at the expense of significantly lower throughput. We overcome this limitation by developing \textsf{BOA} and \textsf{BOA+} that perform online adaptation of beamwidth to minimize wasteful computation and maximize throughput. Our experiments show that more challenging the workload, the greater are the improvements in throughput.

%

\end{document}